\documentclass[conference]{IEEEtran}
\usepackage[acronyms,shortcuts]{glossaries}
\usepackage{cite}
\usepackage{amsmath,amssymb,amsfonts}
\usepackage{algorithmic}
\usepackage{textcomp}

\usepackage{algorithmic}            
\usepackage{algorithm}
\usepackage{amsmath,amssymb,amsfonts,wasysym,latexsym}
\usepackage{graphicx,caption,subcaption,xspace,xcolor,cite}
\usepackage{xcolor}

\def\BibTeX{{\rm B\kern-.05em{\sc i\kern-.025em b}\kern-.08em
    T\kern-.1667em\lower.7ex\hbox{E}\kern-.125emX}}
\newcommand{\argmin}{\mathop{\rm argmin}}
\newcommand{\normof}[2]{\left\|#1\right\|_{#2}}
\newcommand{\fronorm}[1]{\normof{#1}{\rm F}} 

\newcommand{\nmode}[2]{\big[\boldsymbol{{\mathcal{#1}}}\big]_{\left(#2\right)}}
\newcommand{\bb}[1]{\mathbb{#1}}
\newcommand{\ma}[1]{\boldsymbol{#1}}
\newcommand{\unvec}[1]{\text{unvec}\big( #1 \big)}
\renewcommand{\vec}[1]{\text{vec}\big( #1\big)}

\def\BibTeX{{\rm B\kern-.05em{\sc i\kern-.025em b}\kern-.08em
    T\kern-.1667em\lower.7ex\hbox{E}\kern-.125emX}}

\newacronym{2G}{2G}{second generation}
\newacronym{3G}{3G}{third generation}
\newacronym{4G}{4G}{fourth generation}
\newacronym{5G}{5G}{fifth generation}
\newacronym{B5G}{B5G}{beyond fifth generation}
\newacronym{6G}{6G}{sixth generation}
\newacronym{3GPP}{3GPP}{3$\text{rd}$~Generation Partnership Project}
\newacronym{LTE}{LTE}{long term evolution}
\newacronym{NR}{NR}{new radio}
\newacronym{LS}{LS}{least squares}

\newacronym{IRS}{IRS}{intelligent reconfigurable surface}
\newacronym{RIS}{RIS}{reconfigurable intelligent surface}
\newacronym{LIS}{LIS}{large intelligent surface}
\newacronym{SDS}{SDS}{software-defined surface}

\newacronym{D2D}{D2D}{device-to-device}
\newacronym{BS}{BS}{base station}
\newacronym{UE}{UE}{user equipment}

\newacronym{SU}{SU}{single-user}
\newacronym{MU}{MU}{multi-user}
\newacronym{SISO}{SISO}{single-input single-output}
\newacronym{MISO}{MISO}{multiple-input single-output}
\newacronym{SIMO}{SIMO}{single-input multiple-output}
\newacronym{MIMO}{MIMO}{multiple-input multiple-output}

\newacronym{CSI}{CSI}{channel state information}
\newacronym{LOS}{LOS}{line-of-sight}
\newacronym{NLOS}{NLOS}{non-line-of-sight}

\newacronym{QoS}{QoS}{quality-of-service}
\newacronym{SE}{SE}{spectral efficiency}
\newacronym{EE}{EE}{energy efficiency}
\newacronym{SINR}{SINR}{signal to interference plus noise ratio}
\newacronym{SNR}{SNR}{signal to noise ratio}

\newacronym{ProSe}{ProSe}{proximity services}
\newacronym{NSPS}{NSPS}{national security and public safety}

\newacronym{RRM}{RRM}{radio resource management}
\newacronym{MS}{MS}{mode selection}
\newacronym{RA}{RA}{resource allocation}
\newacronym{PC}{PC}{power control}

\newacronym{BCD}{BCD}{block coordinate descent}

\newacronym{RF}{RF}{radio frequency}
\newacronym{AWGN}{AWGN}{additive white Gaussian noise}
\newacronym{MRC}{MRC}{maximum ratio combining}

\newacronym{AF}{AF}{amplify-and-forward}
\newacronym{DF}{DF}{decode-and-forward}
\newacronym{DFT}{DFT}{discrete Fourier transform}
\newacronym{TX}{TX}{transmitter}
\newacronym{RX}{RX}{receiver}
\newacronym{ALS}{ALS}{alternating least squares}
\newacronym{BALS}{BALS}{bilinear alternating least squares}
\newacronym{SVD}{SVD}{singular value decomposition}
\newacronym{HOSVD}{HOSVD}{high order singular value decomposition}
\newacronym{THOSVD}{THOSVD}{truncated high order singular value decomposition}
\newacronym{PARAFAC}{PARAFAC}{PARAllel FACtors}
\newacronym{AOD}{AOD}{angle of departure}
\newacronym{AOA}{AOA}{angle of arrival}
\newacronym{URA}{URA}{uniform rectangular array} 
\newacronym{ADR}{ADR}{achievable data rate}
\newacronym{NMSE}{NMSE}{normalized mean square error}
\newacronym{SER}{SER}{symbol error rate}
\newacronym{LRA}{LRA}{low-rank approximation}

\newacronym{ULA}{ULA}{uniform linear array}
\newacronym{mmWave}{mmWave}{milimiter-wave}
\newacronym{CS}{CS}{compressed sensing}
\newacronym{OFDM}{OFDM}{orthogonal frequency division multiplexing}
\newacronym{PIN}{PIN}{positive-intrinsic-negative}
\newacronym{BD-RIS}{BD-RIS}{beyond diagonal reconfigurable intelligent surface}
\newacronym{LS-Kron}{LS-Kron}{least squares Kronecker factorization}

\newacronym{BTALS}{BTALS}{block Tucker alternating least squares}
\newacronym{BTKF}{BTKF}{block Tucker Kronecker factorization}
\newacronym{PALS}{PALS}{PARAFAC alternating least squares}
\newacronym{PKF}{PKF}{PARAFAC Khatri-Rao factorization}
\newacronym{LS-KRF}{LS-KRF}{Least-Squares Khatri-Rao Factorization}
\newacronym{KRF}{KRF}{Khatri-Rao Factorization}

\IEEEoverridecommandlockouts 

\begin{document}

\title{A Decoupled Channel Estimation Method for Beyond Diagonal RIS\\
}


 \author{\IEEEauthorblockN{Bruno Sokal, Fazal-E-Asim, André L. F. de Almeida, Hongyu Li, and Bruno Clerckx}
 }

\maketitle

\begin{abstract}
\Ac{BD-RIS} is a new architecture for RIS where elements are interconnected to provide more wave manipulation flexibility than traditional single-connected RIS, enhancing data rate and coverage. However, channel estimation for \ac{BD-RIS} is challenging due to the more complex multiple-connection structure involving their scattering elements. To address this issue, this paper proposes a decoupled channel estimation method for BD-RIS that yields separate estimates of the involved channels to enhance the accuracy of the overall combined channel by capitalizing on its Kronecker structure. Starting from a least squares estimate of the combined channel and by properly reshaping the resulting filtered signal, the proposed algorithm resorts to a \acf{KRF} method that teases out the individual channels based on simple rank-one matrix approximation steps. Numerical results show that the proposed decoupled channel estimation yields more accurate channel estimates than the classical least squares scheme. 
\end{abstract}


\section{Introduction}\label{Sec:Introduction}

A recent advancement over the conventional \acf{RIS} technology, which uses diagonal phase shift matrices \cite{basar2019,jian2022reconfigurable}, is the \acf{BD-RIS} which proved to achieve enhanced channel gain and enlarged coverage \cite{Clerckx_Shanpu_BD_RIS_network,Clerckx_CM_MAR_2024,CLX_BD_GC_2023}. Interconnecting elements via additional tunable components enables BD-RIS to mathematically generate scattering matrices with nonzero off-diagonal entries, increasing the flexibility to manipulate waves. It is worth noting that the enhanced performance of BD-RIS architectures and modes depends highly on the accuracy of the channel state information (CSI). However, it is difficult to effectively and efficiently acquire the CSI for BD-RIS-aided wireless systems due to the lack of signal processing capabilities, from which, as the conventional RIS,  the combined TX-RIS-RX channel is often estimated \cite{swindlehurst2022channel,zheng2022survey,gil2021}. Also, the BD-RIS has additional physical constraints due to the physical connections, imposing a challenge on the design of the training BD-RIS matrix. 

For conventional RIS-assisted systems, a traditional \ac{LS} based method has been proposed for combined TX-RIS-RX channel estimation. Although \ac{LS} methods demonstrate good performance in terms of \ac{NMSE}, LS methods often suffer from long pilot overhead and do not exploit the intrinsic structure of the combined TX-RIS-RX channel. In \cite{AraujoSAM2020,gil2021}, the authors proposed a decoupled channel estimation method that exploits the structure of the combined TX-RIS-RX channel. Their proposed method relies on tensor-based signal processing due to the ability of tensor tools to fully exploit different signal diversity \emph{dimensions} (e.g., space, time, frequency, polarization, etc.) \cite{de2007parafac_Andre,confac,FavierAlmeida2014,Almeida2014STF}. In this sense, several works have adopted tensor-based methods for channel estimation \cite{Alexandropoulos_2020,ardah2021trice,benicio2023tensor_wcl}. In \cite{Alexandropoulos_2020}, the authors proposed a tensor-based approach for RIS-assisted multi-user MISO systems. The work \cite{ardah2021trice} proposed a tensor-based framework for parametric estimation of low-rank channels. In \cite{benicio2023tensor_wcl}, the authors proposed a tensor-based method for channel estimation and data in time-varying channels.  Also, tensor-based methods can be employed in different applications in RIS-assisted systems, e.g., in  \cite{Sokal_2023}, the authors introduced a tensor-based method that decouples the optimum phase shift vector of the RIS into multiple smaller factors, reducing the total required control signaling overhead between the network control and the RIS.

For BD-RIS, few works in the literature addressed the channel estimation problem \cite{Clerckx_Arxiv_2024,Sokal_BD_ris}. The work \cite{Clerckx_Arxiv_2024} has proposed a baseline \ac{LS} based receiver to estimate the combined channel based on an orthogonal BD-RIS training matrix design that fulfills the BD-RIS physical constraints. In \cite{Sokal_BD_ris}, the authors proposed a tensor-based approach for channel estimation and design of the BD-RIS training matrix, where an iterative and a closed-form-based receiver is proposed.


In this paper, we formulate a decoupled channel estimation method for MIMO BD-RIS systems. We consider the same system aspect and BD-RIS training design as the one proposed in the baseline LS solution \cite{Clerckx_Arxiv_2024}. Starting from the LS estimate of the combined channel and by exploiting its Krenecker decomposition structure, we resort to a \acf{KRF} method that reveals the individual channels based on simple rank-one matrix approximation steps. The proposed method is conceptually simple and can be executed at low complexity since it relies on \emph{permute} and \emph{reshaping} operations on the filtered signal followed by multiple independent rank-one approximations. Interestingly, it is built up on the same principle as the \acf{KRF} algorithm proposed earlier in \cite{gil2021} for traditional single-connected RIS. Otherwise stated, the proposed decoupled BD-RIS channel estimation method can be viewed as a general approach that encompasses the single-connected RIS to the group and fully-connected RIS (BD-RIS). Numerical results corroborate the effectiveness of the proposed method to provide accurate and individual channel estimates for BD-RIS.

\section{Notation and properties}\label{Sec:notation}

Scalars are represented as non-bold lower-case letters $a$, column vectors as lower-case boldface letters $\ma{a}$, matrices as upper-case boldface letters $\ma{A}$. The superscripts $\{\cdot\}^{\text{T}}$, $\{\cdot\}^{\text{*}}$, $\{\cdot\}^{\text{H}}$  stand for transpose, conjugate, conjugate transpose. \textcolor{black}{An identity matrix of dimension $K$ is denoted as $\ma{I}_{K}$.} The operator $\Arrowvert\cdot\Arrowvert_{\text{F}}$ denotes the Frobenius norm of a matrix or tensor, and $\bb{E}\{\cdot\}$ is the expectation operator. The operator $\text{diag}\left(\ma{a}\right)$ converts $\ma{a}$ into a diagonal matrix, while $\text{diag}(\ma{A})$ returns a vector whose elements are the main the diagonal of $\ma{A}$. \textcolor{black}{Given a matrix $\ma{A} \in \bb{C}^{I \times R}$, the operator $\text{D}_{i}(\ma{A})$ defines a diagonal matrix of size $R \times R$ constructed from the $i$-th row of $\ma{A}$, for $i \in \{1,\ldots,I\}$.} From a set of $Q$ matrices $\ma{X}^{(q)} \in \bb{C}^{M \times N}$, $q = \{1,\ldots,Q\}$, we can construct a block diagonal matrix as \textcolor{black}{ $\ma{X} = \text{bdiag}\big(\ma{X}^{(1)},\ldots,\ma{X}^{(Q)}\big)  \in \bb{C}^{MQ \times NQ}$}. Moreover, $\text{vec}\left(\ma{A}\right)$ converts $\ma{A} \in \mathbb{C}^{I \times R}$ to a column vector $\ma{a} \in \mathbb{C}^{IR \times 1}$ by stacking its columns on top of each other, while the operator \textcolor{black}{$\unvec{\ma{a}}_{I \times R}$} returns to the matrix $\ma{A} \in \bb{C}^{I \times R}$. The symbols $\circ$, $\otimes$, and $\diamond$ denote the outer product, the Kronecker product, and the Khatri-Rao product (also known as the column-wise Kronecker product), respectively. For example, the Khatri-Rao product of matrices $\ma{X} \in \bb{C}^{I \times R}$ and $\ma{Y} \in \bb{C}^{J \times R}$, is defined as 
\begin{equation}
  \label{eq:khatri_rao_def}  \ma{Z} = \ma{X} \diamond \ma{Y} = [\ma{x}_1 \otimes \ma{y}_1, \ldots,\ma{x}_R  \otimes \ma{y}_R] \in \bb{C}^{JI \times R},
\end{equation}
where $\ma{x}_r$ and $\ma{y}_r$ are the $r$-th column of $\ma{X}$ and $\ma{Y}$, respectively, for $r=\{1,\ldots,R\}$.
Useful properties that will be exploited in this paper are
\begin{align}
\label{prop:vec}\text{vec}\left(\ma{A}\ma{B}\ma{C}\right) &= \left(\ma{C}^{\text{T}} \otimes \ma{A}\right)\text{vec}\left(\ma{B}\right), \\
\label{prop:vec_diag}  \text{vec}\left(\ma{A}\text{diag}\left( \ma{b}\right)\ma{C}\right)  &= \left(\ma{C}^{\text{T}} \diamond \ma{A}\right)\ma{b},\\
\label{prop:vec_permute}\text{vec}(\ma{A} \otimes \ma{B}) &= \ma{P}\big(\text{vec}(\ma{A}) \otimes \text{vec}(\ma{B})\big) \\
 \label{prop:vec_r1} \text{vec}\left(\ma{a}\ma{b}^{\text{T}}\right) &= \ma{b} \otimes \ma{a} 
\end{align}
where the involved vectors and matrices have compatible dimensions in each case, and $\ma{P}$ is a permutation matrix.

  \begin{figure}[!t]
\centering
  \includegraphics[width=0.9\linewidth]{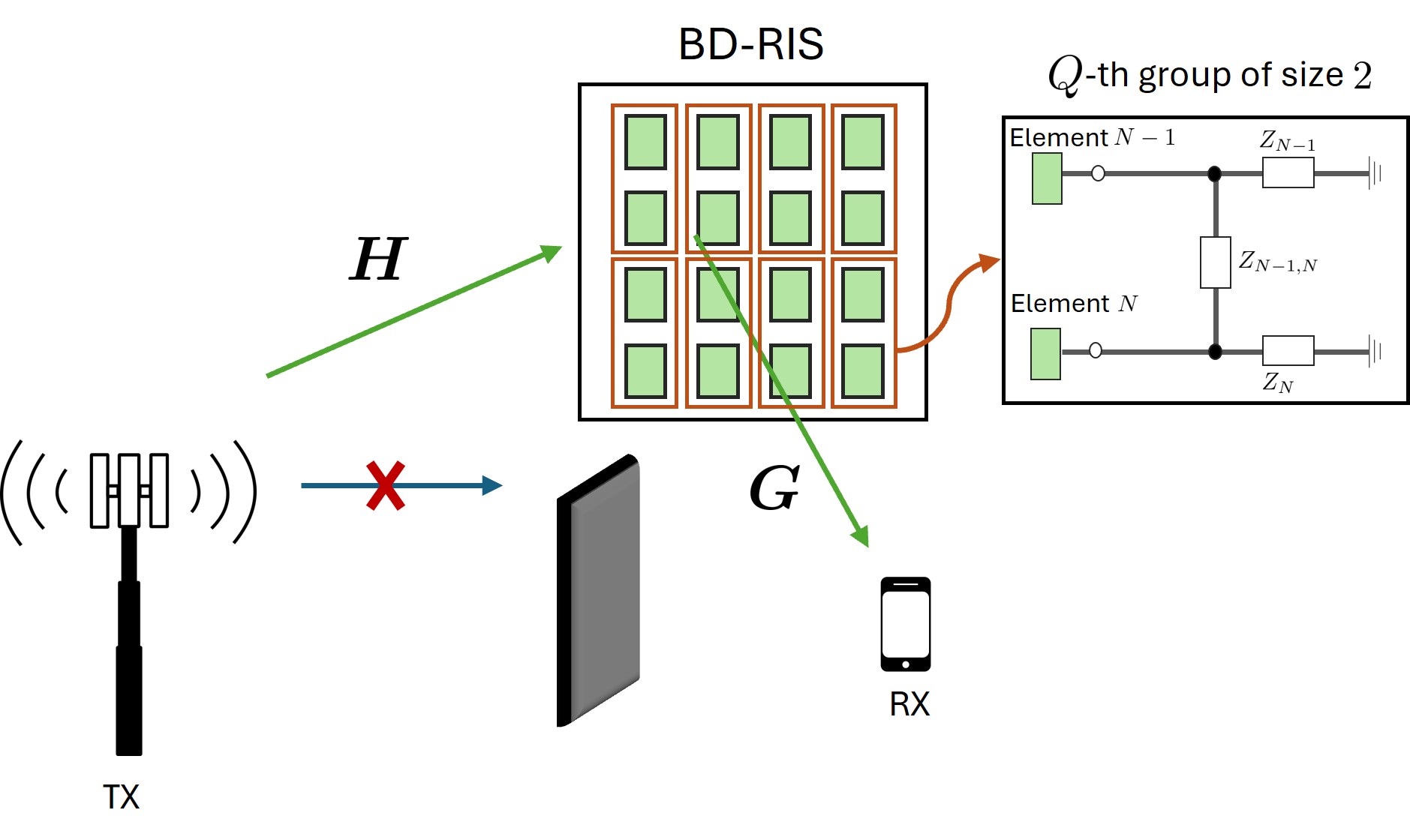}
  \caption{BD-RIS assisted communication scenario.}
  \label{fig:system_model}

\end{figure}

\section{System Model}\label{Sec:System_Model}
We consider a \ac{MIMO} system assisted by a \ac{BD-RIS},  where the transmitter and the receiver are equipped with $M_T$ and $M_R$ antennas, respectively, and a \ac{BD-RIS} with $N$ elements, illustrated in Figure \ref{fig:system_model}. We assume the direct link between the transmitter and the receiver is blocked. Assuming a length-$T$ pilot sequence, the received signal at the $t$ time slot is given by $\ma{y}_t = \ma{G}\ma{S}_{t}\ma{H}^{\text{T}}\ma{x}_{t} + \ma{b}_{t} \in \bb{C}^{M_R \times 1}$, where $\ma{G} \in \bb{C}^{M_R \times N}$  and $\ma{H} \in \bb{C}^{M_T \times N }$ are the IRS-RX and the TX-IRS channels respectively, and $\ma{b}_t$ is the \ac{AWGN} term, with zero mean and unit variance. $\ma{S}_t \in \bb{C}^{N \times N}$ is the BD-RIS scattering matrix, with $\ma{S}^{\text{H}}_{t}\ma{S}_{t} = \ma{I}_N$. Assuming a group connected BD-RIS architecture  \cite{CLX_BD_GC_2023}, 
the $N$ elements are divided into $Q$ groups with size $\bar{N}$, i.e., $N = \bar{N}Q$. In this case, the scattering matrix can be written as $\ma{S}_t = \text{bdiag}(\ma{S}^{(1)}_{t}, \ldots \ma{S}^{(Q)}_{t}) \in \bb{C}^{N \times N}$. Due to the physical constraints of the BD-RIS group connections, the $q$-th scattering matrix $\ma{S}^{(q)}_{t}$ must satisfy $\ma{S}^{(q)\text{H}}_t\ma{S}_t^{(q)} = \ma{I}_{\bar{N}}$. Thus, the received signal can be written as 
\begin{align}
    \ma{y}_t = \sum_{q=1}^Q \ma{G}^{(q)}\ma{S}^{(q)}_{t}\ma{H}^{(q)\text{T}}\ma{x}_t + \ma{b}_t,
\end{align} where $\ma{H}^{(q)} \in \bb{C}^{M_T \times \bar{N}}$ and $\ma{G}^{(q)}\in \bb{C}^{M_R \times \bar{N}}$ correspond to the $q$-th block of  $\ma{H}\in \bb{C}^{N \times \bar{N}Q}$ and  $\ma{G} \in \bb{C}^{M_R \times \bar{N}Q}$, respectively, defined as follows
\begin{align}
   \label{eq:channels}  \ma{H}^{(q)} &= \ma{H}_{.\, [(q-1)\bar{N}+1,\ldots, q\bar{N}]} \in \bb{C}^{M_T \times \bar{N}}, \,\, q = 1,\ldots,Q\\ 
    \ma{G}^{(q)} &= \ma{G}_{.\, [(q-1)\bar{N}+1,\ldots, q\bar{N}]} \in \bb{C}^{M_R \times \bar{N}},\,\, q =1,\ldots,Q
\end{align}
Hence, the channel matrices can be seen as a concatenation of smaller submatrices such that $\ma{H}=[\ma{H}^{(1)},\ldots, \ma{H}^{(Q)}] \in \bb{C}^{M_T \times  \bar{N}Q }$ and $\ma{G}=[\ma{G}^{(1)},\ldots, \ma{G}^{(Q)}] \in \bb{C}^{M_R \times \bar{N}Q}$. Using the property (\ref{prop:vec}), the noise-free received signal can be written as
\begin{align}
   \ma{y}_t \vspace{-0.15cm} &= \vspace{-0.15cm}\sum_{q=1}^{Q} (\ma{x}_t^{\text{T}} \otimes \ma{I}_{M_R})\vec{\ma{G}^{(q)}\ma{S}^{(q)}_{t}\ma{H}^{(q)\text{T}}}   \vspace{-0.15cm} \\
   &=\sum_{q=1}^{Q} (\ma{x}_t^{\text{T}} \otimes \ma{I}_{M_R})(\ma{H}^{(q)} \otimes \ma{G}^{(q)})\vec{\ma{S}^{(q)}_{t}} \\
   &=\sum_{q=1}^{Q}(\vec{\ma{S}^{(q)}_{t}}^{\text{T}} \otimes \ma{x}^{\text{T}}_t \otimes \ma{I}_{M_R})\ma{c}^{(q)}, 
\end{align}
where $\ma{c}^{(q)} = \vec{\ma{H}^{(q)} \otimes \ma{G}^{(q)}}  \in \bb{C}^{M_RM_T\bar{N}^2 \times 1}$. By collecting the $T$ time slots and removing the sum, we obtain
\begin{align}
    \ma{y} &= \big[\ma{y}_1^{\text{T}},\ldots,\ma{y}_T^{\text{T}}\big]^{\text{T}} 
    =\Big[\big(\overline{\ma{S}} \diamond \ma{X}\big)^{\text{T}} \otimes \ma{I}_{M_R}\Big]\ma{c} \\ &=\big[\ma{\Omega} \otimes \ma{I}_{M_R}\big]\ma{c} \in \bb{C}^{M_RT \times 1},
\end{align}
where $\ma{\Omega} =  \big(\overline{\ma{S}} \diamond \ma{X}\big)^{\text{T}} \in \bb{C}^{T \times M_T\bar{N}^2Q}$ is the combined BD-RIS pilot matrix, which is known at the receiver, with  $\overline{\ma{S}} = \big[\ma{s}_1, \ldots, \ma{s}_T\big] \in \bb{C}^{\bar{N}^2Q \times T}$, $\ma{s}_t = \big[\vec{\ma{S}}^{(1)\text{T}}_{t},\ldots,\vec{\ma{S}}^{(Q)^\text{T}}_{t}\big]^{\text{T}} \in \bb{C}^{\bar{N}^2Q \times 1}$,  $\ma{X} = [\ma{x}_1, \ldots \ma{x}_T] \in \bb{C}^{M_T \times T}$ is the pilot matrix, and $\ma{c} = \big[\ma{c}^{(1)\text{T}},\ldots,\ma{c}^{(Q)\text{T}}\big]^{\text{T}} \in \bb{C}^{M_RM_T \bar{N}^{2}Q \times 1}$ is the combined channel, for $q=1,\ldots,Q$. 
The classical approach is to estimate the combined channel $\ma{c} \in \bb{C}^{M_RM_T\bar{N}^2Q\times1}$ resorts to a \acf{LS} method \cite{Clerckx_Arxiv_2024}. This problem can be formulated as
\begin{align}
  \label{eq:ls_problem}  \hat{\ma{c}} &= \underset{\ma{c}}{\argmin} \big\| \ma{y} - (\ma{\Omega}\otimes\ma{I}_{M_R})\ma{c}\big\|^2, 
\end{align}
from which the matching filter gives its solution as
\begin{align}
   \label{eq:comb_channel_sol} \hat{\ma{c}} &= \frac{\bar{N}}{T}(\ma{\Omega}\otimes\ma{I}_{M_R})^{\text{H}}\ma{y}  \\ \notag &\approx \Big[\vec{\ma{H}^{(1)} \otimes \ma{G}^{(1)}}^{\text{T}},\ldots,\vec{\ma{H}^{(Q)} \otimes \ma{G}^{(Q)}}^{\text{T}}\Big]^{\text{T}}  
\end{align}

 This solution requires $T \geq M_T \bar{N}^2Q$ to ensure a unique combined channel estimate. Next, we present the proposed solution for decoupling the estimates of $\ma{G}$ and $\ma{H}$ from the combined channel estimation $\hat{\ma{c}}$.
 
\section{Decoupled channel estimation method}\label{Sec:CE_method}
In this section, based on the LS estimate in (\ref{eq:comb_channel_sol}), we show how to obtain the individual channel estimates.  

\begin{figure}[!t]
\centering
  \includegraphics[width=0.995\linewidth]{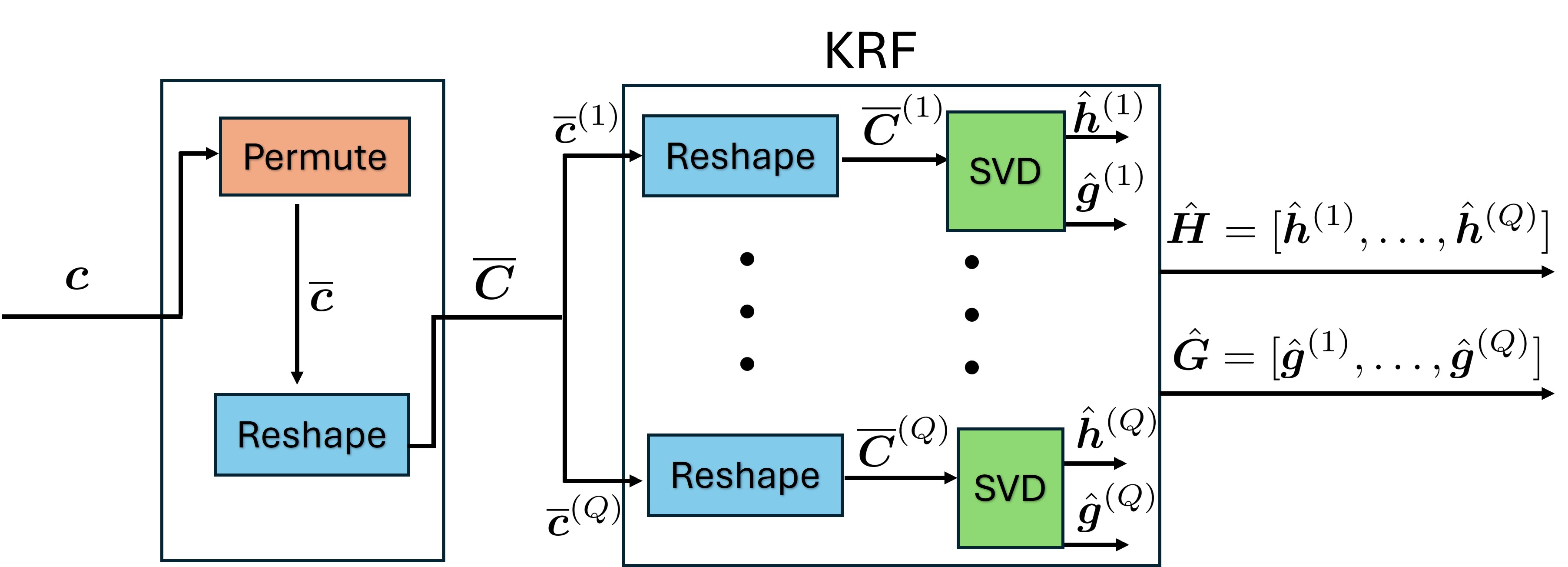}
  \caption{Block diagram of the proposed KRF method.}
  \label{fig:krf_block}
\end{figure}

Let us define the estimated combined channel matrix  $\hat{\ma{C}}$ as
\begin{align}
  \hat{\ma{C}}  &= \unvec{\hat{\ma{c}}}_{M_RM_T\bar{N}^2 \times Q} \\ &\approx  \big[\vec{\ma{H}^{(1)} \otimes \ma{G}^{(1)}},\ldots,\vec{\ma{H}^{(Q)} \otimes \ma{G}^{(Q)}}\big].
\end{align}
The $q$-th column of $\hat{\ma{C}}$ is given by $\hat{\ma{c}}^{(q)} \approx \text{vec}(\ma{H}^{(q)} \otimes \ma{G}^{(q)}) \in \bb{C}^{M_RM_T\bar{N}^2 \times 1}$. By applying property (\ref{prop:vec_permute}), we can define the permuted vector $\overline{\ma{c}}^{(q)} = \ma{P}\hat{\ma{c}}^{(q)} \in \bb{C}^{M_R\bar{N}M_T\bar{N} \times 1}$ as
\begin{align}
    \label{eq:rec_vec_sig_permuted}   \overline{\ma{c}}^{(q)} \approx \vec{\ma{H}^{(q)}} \otimes \vec{\ma{G}^{(q)}}, 
\end{align}
where $\ma{P} \in \bb{R}^{M_R\bar{N}M_T\bar{N} \times M_RM_T\bar{N}^2}$ is a permutation matrix.
By collecting the $Q$ group vectors, we have
\begin{align}
  \notag   \overline{\ma{C}} &= [\overline{\ma{c}}^{(1)},\ldots, \overline{\ma{c}}^{(Q)}] \in \bb{C}^{M_R\bar{N}M_T\bar{N}\times Q}, \\
   \notag  &\approx [\ma{h}^{(1)} \otimes \ma{g}^{(1)},\ldots,\ma{h}^{(Q)} \otimes \ma{g}^{(Q)}] \\
 \label{eq:Cbar}   &\approx \overline{\ma{H}} \diamond \overline{\ma{G}},
\end{align}
where $\overline{\ma{H}} = [\ma{h}^{(1)},\ldots, \ma{h}^{(Q)}] \in \bb{C}^{M_T\bar{N} \times Q}$ and $\overline{\ma{G}} = [\ma{g}^{(1)},\ldots, \ma{g}^{(Q)}] \in \bb{C}^{M_R\bar{N} \times Q}$, with $\ma{h}^{(q)}= \vec{\ma{H}^{(q)}} $ and $\ma{g}^{(q)}= \vec{\ma{G}^{(q)}} $ being the vectorized forms of the $q$-th group of the channels $\ma{G}$ and $\ma{H}$, respectively, $\forall q= 1,\ldots,Q$.
From the definition of the Khatri-Rao product in (\ref{eq:khatri_rao_def}) and  making use of (\ref{prop:vec_r1}), the $q$-th column of $\overline{\ma{C}}$, denoted as $\overline{\ma{c}}^{(q)}$, is reshaped into a rank-one matrix, i.e.,
\begin{align}
   \label{eq:r1_channel} \overline{\ma{C}}^{(q)} = \text{unvec}(\overline{\ma{c}}^{(q)})_{M_R\bar{N}\times M_T\bar{N}} \approx \ma{g}^{(q)}\ma{h}^{(q)\text{T}},
\end{align}
To extract the individual channel estimates from (\ref{eq:r1_channel}), any rank-one matrix approximation method can be used, e.g., truncated \ac{SVD} or a power method. By assuming the rank-one approximation $\overline{\ma{C}}^{(q)} \approx \sigma\ma{u}^{(q)}\ma{v}^{(q)\text{H}}$, where $\ma{u}^{(q)}$ and $\ma{v}^{(q)}$ are the dominant left and right singular vectors associated with the largest singular value $\sigma$, we obtain the channel estimates as
\begin{align}
   \label{eq:channel_est} \hat{\ma{g}}^{(q)} = \sqrt{\sigma}\ma{u}^{(q)}, \quad \hat{\ma{h}}^{(q)} = \sqrt{\sigma}\ma{v}^{(q)*}
\end{align}
The estimates of the channel matrices are given by $\hat{\ma{H}}^{(q)} = \unvec{\hat{\ma{h}}^{(q)}}_{M_T \times \bar{N}}$ and $\hat{\ma{G}}^{(q)} = \unvec{\hat{\ma{g}}^{(q)}}_{M_R \times \bar{N}}$. Repeating the process for the $Q$ columns of $\overline{\ma{C}}$ (corresponding to the $Q$ BD-RIS groups), we build the complete channel matrices as $\hat{\ma{H}} = [\hat{\ma{H}}^{(1)},\ldots,\hat{\ma{H}}^{(Q)}] \in \bb{C}^{M_T\times \bar{N}Q}$ and $\hat{\ma{G}} = [\hat{\ma{G}}^{(1)},\ldots,\hat{\ma{G}}^{(Q)}] \in \bb{C}^{M_R\times \bar{N}Q}$.

\begin{algorithm}[!t]
	\begin{algorithmic}[1]
		\caption{Decoupled Channel Estimation based on Least-Squares Khatri-Rao Factorization (LS-KRF)}\label{algorithm_krf}
 \STATE \textbf{Inputs}: Received signal  $\ma{y}$, \ac{BD-RIS} pilot matrix $\ma{\Omega}$
            \STATE Compute the LS estimate via a match filtering
            \begin{equation*}
                \ma{\hat{c}}  \approx \frac{\bar{N}}{T} \big( \ma{\Omega} \otimes \ma{I}_{M_R}\big)^{\text{H}}\ma{y} \in \bb{C}^{M_RM_T\bar{N}^2Q \times 1}
            \end{equation*}
            \STATE Permute the filtered signal: $\overline{\ma{c}} = \ma{P}\ma{\hat{c}} \in \bb{C}^{M_R\bar{N}M_T\bar{N}Q \times 1}$.
            \STATE Obtain $\overline{\ma{C}} = \text{unvec}(\overline{\ma{c}})_{M_R\bar{N}M_T\bar{N} \times Q}  \approx \nmode{H}{3}^{\text{T}} \diamond \nmode{G}{3}^{\text{T}}$.
\FOR{$q = 1:Q$}
\STATE Let $\overline{\ma{c}}^{(q)} \approx \ma{h}^{(q)} \otimes \ma{g}^{(q)}  \in \bb{C}^{M_R\bar{N}M_T\bar{N}M_T \times 1}$ be the $q$-th column of $\overline{\ma{C}}$. Obtain $\overline{\ma{C}}^{(q)}$ as
\begin{equation*}
    \overline{\ma{C}}^{(q)} = \text{unvec}(\overline{\ma{c}}^{(q)})_{M_R\bar{N}\times M_T\bar{N}}  \approx \ma{g}^{(q)} \ma{h}^{(q)\text{T}}
\end{equation*}
 \STATE From the rank-one approximation (truncated SVD) $\overline{\ma{C}}^{(q)}\approx  \sigma^{(q)}\ma{u}^{(q)}\ma{v}^{(q)\text{H}}$, obtain the channel estimates as
\begin{equation*}
    \hat{\ma{g}}^{(q)} = \sqrt{\sigma}\ma{u}^{(q)}, \, \quad \hat{\ma{h}}^{(q)} = \sqrt{\sigma}\ma{v}^{(q)*}
\end{equation*}
\STATE Define the $q$-th group channel matrices as
\begin{equation*}
    \hat{\ma{G}}^{(q)} = \text{unvec}(\hat{\ma{g}}^{(q)})_{M_R \times \bar{N}}, \, \,
    \hat{\ma{H}}^{(q)} = \text{unvec}(\hat{\ma{h}}^{(q)})_{M_T \times \bar{N}}. 
\end{equation*}
 \ENDFOR
 \RETURN{}  $\hat{\ma{G}} = [\hat{\ma{G}}^{(1)},\ldots,\hat{\ma{G}}^{(Q)}]$ and $\hat{\ma{H}} = [\hat{\ma{H}}^{(1)},\ldots,\hat{\ma{H}}^{(Q)}]$   
	\end{algorithmic}
\end{algorithm}

\section{Computational Complexity}\label{Sec:Comp_Complex}
The proposed KRF decoupled channel estimation method is an additional step to the LS combined channel estimator. By considering the matching filtering in (\ref{eq:comb_channel_sol}), we have a computational cost of $\mathcal{O}(M_RT^2)$, where $T = T_\text{min} = M_T\bar{N}^2Q$ is assumed. The KRF method requires computing $Q$ rank-one approximations of an $M_R\bar{N} \times M_T\bar{N}$ matrix. Assuming that for a matrix $\ma{A}$ of size $m \times n$, the total complexity for its rank-one approximation is given by $\mathcal{O}(mn)$ \cite{flops_rank1}, the KRF method complexity is given by $\mathcal{O}(M_RT)$. Consequently, the computational complexity of the KRF method is given by $\mathcal{O}(M_RT^2 + M_RT)$. Note that the overall gap, in terms of computational complexity, between the LS and the KRF method becomes negligible since the bulk of the complexity is associated with the LS filtering stage. Also, notice that the $Q$ rank-one approximation steps used in KRF (see Figure \ref{fig:krf_block}) can be executed in parallel, minimizing the processing delay associated with the decoupled channel estimation step.

\begin{figure}[!t]
\centering
  \includegraphics[width=0.86\linewidth]{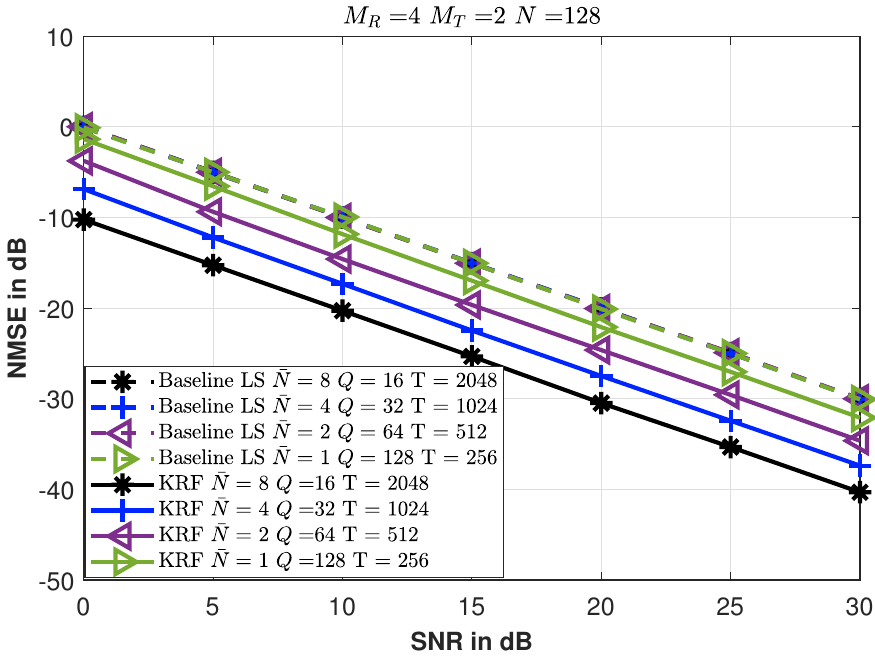}
  \caption{NMSE performance varying the total pilot overhead $T$.}
  \label{fig:NMSE_SNR_vary_T}
  \end{figure}
\begin{figure}[!t]
\centering
  \includegraphics[width=0.86\linewidth]{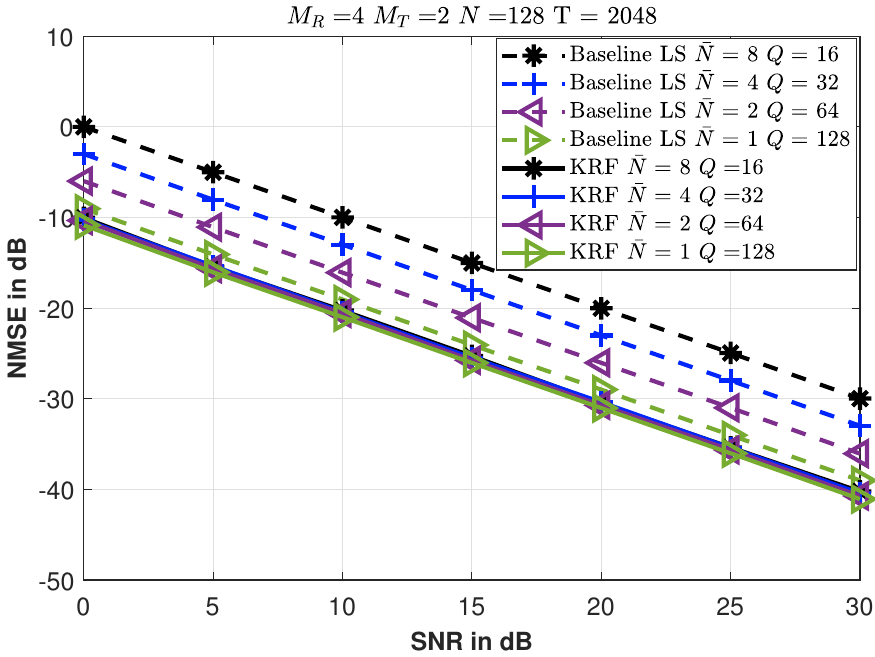}
  \caption{NMSE performance by fixing the  pilot overhead $T$.}
  \label{fig:NMSE_SNR_fix_T}
  \end{figure}

\section{Simulation results}\label{Sec:Simulation_Results}
In this section, we evaluate the performance of the proposed KRF method, comparing it with the baseline LS-based method \cite{Clerckx_Arxiv_2024} in terms of \ac{NMSE} of the combined channel $\ma{C} \in \bb{C}^{M_RM_T \times \bar{N}^2Q}$, given by 
\begin{equation*}
    \text{NMSE} = \frac{1}{L} \sum_{l=1}^{L} \frac{\fronorm{\ma{C}_{(l)} - \hat{\ma{C}}_{(l)}}^2}{\fronorm{\ma{C}_{(l)}}^2}
\end{equation*} with $L=100$ being the number of trials. For the pilot sequence, we assume an orthogonal design by using a truncated \ac{DFT} structure for $\ma{X} \in \bb{C}^{M_T \times T}$, such that $\ma{X}^{\text{H}}\ma{X} = T \ma{I}_{M_T}$. Also, we adopt the orthogonal design of \cite{Clerckx_Arxiv_2024} for the BD-RIS training matrix $\ma{S}$.

Since the pilot overhead is given by $T = M_T \bar{N}^2Q = M_T \bar{N}N$, the group size has the major impact. Thus, depending on the group size $\bar{N}$, the pilot overhead changes even for a fixed number of RIS elements $N$. In Figures \ref{fig:NMSE_SNR_vary_T} and \ref{fig:NMSE_SNR_fix_T}, a RIS panel with $N = 128$ elements is assumed. Starting with Figure \ref{fig:NMSE_SNR_vary_T}, we evaluate the NMSE performance for different group sizes by setting the pilot sequence length to its minimum feasible value for each configuration, i.e., $T_{\text{min}}=M_T\bar{N}^2Q$. Note that the performance of the baseline LS method is insensitive to an increase in the group size (and thus pilot overhead) since it is based on the estimation of the combined channel only, whose number of channel coefficients ($\bar{N}^2Q$) remains unchanged regardless of the group size. However, when including the KRF step to decouple the channel matrices, we observe that the performance increases with the group size $\bar{N}$. This comes from the fact that the proposed method exploits the Kronecker structure of the combined channel in (\ref{eq:ls_problem}) after the LS filtering to decouple the channel matrices. This procedure yields refined channel estimates due to noise rejection obtained from the rank-one factorizations that are inherent to the KRF algorithm. In the end, this results in an improved accuracy of the reconstructed combined channel $\hat{\ma{C}}$. 

In Figure \ref{fig:NMSE_SNR_fix_T}, we fix the training overhead to $T=2048$ (the minimum value for the configuration with $\bar{N}=8$). We can observe that the baseline LS method offers an improved performance as the group size $\bar{N}$ is reduced. Indeed, for a fixed training overhead $T$, when the group size $\bar{N}$ becomes smaller, the total number of channel coefficients is decreased accordingly. Therefore, we can conclude that the single-connected architecture ($\bar{N}=1$) is preferable for channel estimation when the LS method is used. On the other hand, the proposed KRF method exhibits approximately the same performance for all the considered group sizes. This is due to the fact that our channel decoupling method refines the combined channel estimate at the output of the LS stage by extracting the individual estimates of the involved channels through multiple rank-one factorizations. This procedure benefits from a noise rejection gain that increases when the group size $\bar{N}$ becomes larger \cite{gil2021}.  In addition, as shown in Fig. \ref{fig:NMSE_SNR_fix_T}, smaller group sizes $\bar{N}$ are compensated by larger training lengths $T$, yielding approximately the same performance for all configurations.

\begin{figure}[!t]
\centering
  \includegraphics[width=0.86\linewidth]{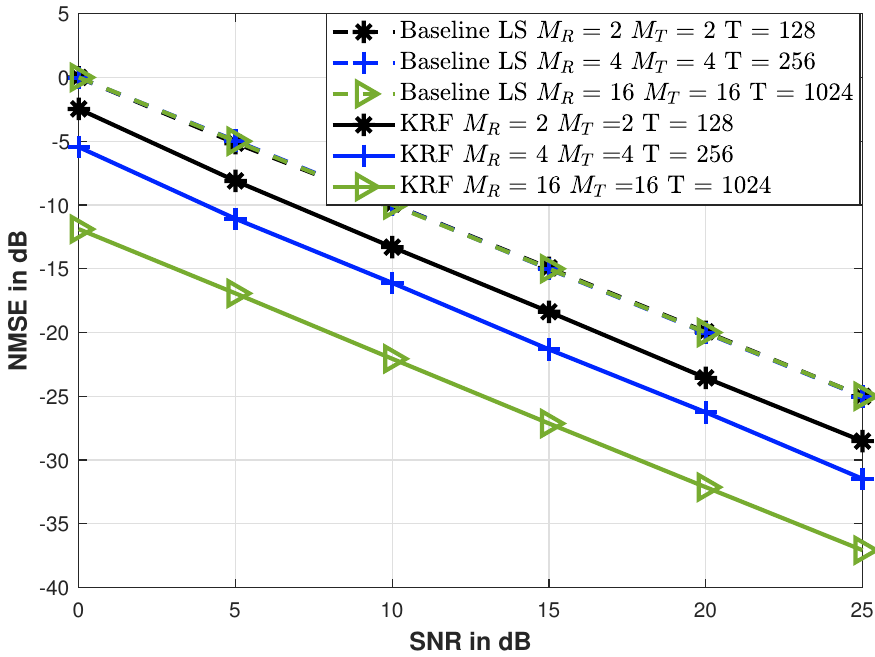}
  \caption{NMSE performance for $N=32$, $\bar{N}=2$, and $Q=16$.}
  \label{fig:NMSE_SNR_vary_T_Mt}

\end{figure}

In Fig. \ref{fig:NMSE_SNR_vary_T_Mt}, we observe that the LS keeps approximately a constant gain by varying the numbers $M_R$ and $M_T$ of receive and transmit antennas, while the KRF performance is enhanced when $M_R$ and/or $M_T$ increases, being able to extract more accurate channel estimates as higher levels of transmit/receive spatial diversities are available. This remarkable result corroborates the effectiveness of the decoupled BD-RIS channel estimation, which benefits from noise rejection gains provided by the rank-one approximation steps of the KRF algorithm.

\begin{figure}[!t]
\centering
  \includegraphics[width=0.86\linewidth]{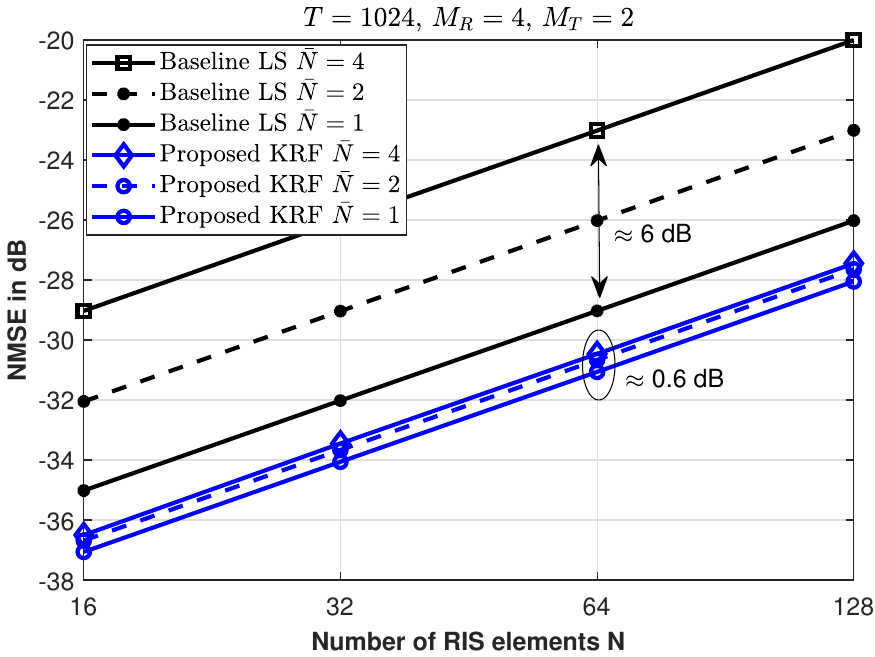}
  \caption{NMSE  varying the number of RIS elements.}
  \label{fig:NMSE_SNR_vary_N}

\end{figure}

In Fig. \ref{fig:NMSE_SNR_vary_N}, the performances are compared for a fixed SNR of $20$ dB and a fixed training overhead $T= 1024$. Here, we point out that by varying the number of BD-RIS elements $N = \bar{N}Q$, the performance gap of the proposed method between single-connected architecture ($\bar{N}=1$) and the group-connected architecture ($\bar{N}>1$) is negligible when the proposed method is used. In contrast, the NMSE performance of the LS receiver significantly depends on the connection degree of the BD-RIS, exhibiting performance gaps between the single-connected and the group-connected architectures.

\section{Conclusions}\label{Sec:Conclusions}
We have proposed a decoupled channel estimation method for \ac{BD-RIS} assisted systems. Our method efficiently exploits the intrinsic Kronecker structure of the combined channel by performing simple permute and reshape operations to filtered signals, allowing the extraction of individual channel estimates via multiple independent rank-one approximations. As shown in our numerical results, decoupling the combined channel estimate into its constituent individual components provides remarkable gains over the baseline LS estimation scheme while offering similar performances regardless of the connection degree of the BD-RIS for fixed training overheads.

\end{document}